# Dynamic learning of synchronization in coupled nonlinear systems

Yong Wu, Qianming Ding, Weifang Huang, Tianyu Li, Dong Yu, Ya Jia*
*Department of Physics, Central China Normal University, Wuhan 430079, China*

**Abstract**：Synchronization phenomena are pervasive in coupled nonlinear systems across the natural world and engineering domains. Understanding how to dynamically identify the parameter space (or network structure) of coupled nonlinear systems in a synchronized state is crucial for the study of system synchronization. To address the challenge of achieving stable synchronization in coupled nonlinear systems, we develop a set of mathematical optimization techniques for dynamic learning of synchronization (DLS) inspired by machine learning. This technology captures the state differences between nodes within the system and dynamically adjusts weights, allowing coupled nonlinear systems to maintain a stable state of synchronization after appropriate weight adjustments. To enhance synchronization optimization, we use the Master Stability Function (MSF) to demonstrate how DLS effectively adjusts networks into their synchronization regions. We introduce several variants of the DLS technique, including adaptive, supervised, and hybrid methods, effectively promoting synchronization in heterogeneous networks such as small-world, scale-free, and random networks. The efficacy of this technique is validated through its application to simple FitzHugh-Nagumo neural networks and complex Hodgkin-Huxley neuronal networks, examining its impact on both global and local synchronization. The DLS technique proposed in this study offers a new solution to synchronization problems in dynamic network environments, addressing the deficiencies in adaptability and flexibility of existing technologies and providing a fresh perspective for understanding and implementing synchronization phenomena in coupled nonlinear systems.

**Keywords**: dynamic weight adjustment; global and local synchronization; linear and nonlinear coupling; complex network

*Corresponding author: jiay@ccnu.edu.cn

## 1. Introduction

Exploring complex systems, especially those composed of coupled oscillators, is crucial for simulating real-world systems and gaining a deep understanding of natural phenomena [1-3]. Such research typically focuses on the phenomenon of synchronization within complex nonlinear coupled oscillatory systems, where oscillators within the system can achieve a uniform state under specific conditions, known as the synchronization mode [4]. This synchronization phenomenon has a profound impact on the function and performance of coupled nonlinear systems, thus attracting widespread attention in interdisciplinary scientific research [5-7].

In the field of physics, the phenomenon of synchronization in coupled nonlinear systems mainly focuses on the exploration of oscillators and their wave behaviors, particularly the synchronization of bidirectionally coupled chaotic oscillators, revealing the complex interactions within the dynamics of physical systems [8]. In engineering and computational science, the principles of synchronization are widely applied in communication and control systems, such as the use of chaotic synchronization for secure encrypted communications, and the importance of maintaining system stability and efficiency in distributed computing and communication [9,10]. Additionally, in the biological sciences, especially in the study of the nervous system, the phenomenon of synchronization plays a fundamental role in the transmission of information between neurons, providing key insights for understanding neural functions and developing potential medical applications [11].

For research on network synchronization, exploring the impact of external factors such as time delays [12], temperature changes [13], and noise interference [14] on the synchronization process of nonlinear neural network systems is particularly crucial. In chaotic systems, research that introduces various external signals to control synchronization behavior has opened a promising new field, indicating new directions for future research efforts [15-18]. Moreover, studying how to fine-tune the properties of network systems and redistribute oscillators in networks with different connectivity degrees to enhance the synchronization performance of nonlinearly coupled networks shows great potential despite challenges [19-24]. However, these traditional methods mainly face challenges in practice due to their high dependence on extensive prior research and experience [25-27]. Although significant progress has been made in understanding synchronization phenomena, achieving efficient and stable synchronization in dynamically changing network environments remains a challenge. This prompted us to develop the dynamic learning of synchronization (DLS) technique, which dynamically adjusts the weights between network nodes to flexibly adapt to network changes and external disturbances, effectively enhancing synchronization performance. Recently, while machine learning algorithms have made some progress in solving network synchronization issues [28], the unique introduction of DLS technology has demonstrated its significant potential and advantages in dealing with synchronization problems in dynamic network environments. This is especially true for its ability to flexibly adjust the weights between network nodes, further strengthening adaptability to network changes and external disturbances.

Achieving network synchronization is key to adjusting the weights of connections between oscillators. Inspired by how biological neurons adjust connection strengths based on activity, researchers have developed methods such as the Hebb rule [29] and Spike-Timing-Dependent Plasticity (STDP) [30-33] that adjust weights based on oscillator activity. Although effective, these methods often rely on external factors, such as time delays, to regulate the state of synchronization [34-36]. Advances in Artificial Neural Networks (ANN), especially the application of backpropagation for weight updating [37], and the use of algorithms like the FORCE rule [38-40], have facilitated rapid adaptation to external changes, bringing new depth to synchronization research. Against this technical backdrop, this paper introduces the Dynamic Learning Synchronization (DLS) technique, which employs an innovative strategy by monitoring network status in real time and dynamically adjusting connection weights. This approach offers a more adaptable strategy for network changes and external disturbances. The DLS technique significantly reduces the need for external condition adjustments, not only enhancing the efficiency and stability of the synchronization process but also paving new paths for understanding and controlling synchronization phenomena in complex networks.

This study introduces an adaptive synchronization strategy that finely tunes the network synchronization process in dynamically changing environments by incorporating the recursive least squares method, demonstrating higher adaptability and efficiency compared to traditional methods. Compared to the research of Sorrentino and Ott, this strategy significantly enhances the reaction speed and precision of synchronization within constantly changing network structures, making network synchronization less dependent on stable network topology [41]. Additionally, compared to the traditional feedback mechanisms emphasized by Ravoori et al., this strategy significantly reduces dependency on external information, speeding up network response through internal mechanism adjustments [42]. Furthermore, considering the contributions of Zhou and Kurths in the area of network synchronization stability, this study optimizes the synchronization process by more flexibly adjusting network weights and connection strengths, enhancing network robustness in the face of scaling or increased dynamics [43]. Compared to the work of Wang, Sun, and Cao, this research demonstrates higher flexibility and efficiency in adjusting coupling weights, better adapting to the specific needs of different network environments [44].In contrast to the neural network structural plasticity focused on by Solís-Perales and Estrada, the strategy developed in this study is not only applicable to biological models but can also be broadly applied to technical networks, providing a synchronization solution with broad application prospects [45]. The comprehensiveness and flexibility of this approach show significant advantages both theoretically and in practical applications, offering effective tools for dealing with rapidly changing network environments.

In this study, we introduce an innovative method named DLS in section 2, focusing on adjusting the weights between network nodes to achieve network synchronization. This technique uniquely considers the real-time state of the network or external inputs, thereby enabling flexible adaptation to environmental changes. To validate the effectiveness and practicality of the DLS technology, we utilize two classical models: FitzHugh-Nagumo (FHN) [46] and Hodgkin-Huxley (HH) [47] in section 3. These models are coupled through electrical and chemical coupling methods to simulate different types of network environments and synchronization conditions. To quantitatively assess the impact of DLS technology on different network synchronization modes, we monitored the standard deviation changes in neural network nodes before and after the application of DLS. Additionally, we introduced a synchronization factor as an evaluation metric to quantify the improvement in network synchronization performance by DLS technology. Incorporating the master stability function (MSF) has allowed us to refine our approach by analytically determining the stability of synchronization states under varying network conditions, further enhancing our understanding of dynamic network synchronization. Using these two key indicators, we could comprehensively analyze the performance of DLS technology in promoting network synchronization and how it effectively adapts to the dynamic changes in the network and external disturbances in section 4. The research results clearly show that DLS technology can significantly optimize network synchronization modes, not only excelling in enhancing synchronization efficiency but also improving stability. Finally, the study summarizes the research findings on DLS technology and discusses its importance in deeply understanding synchronization phenomena in complex systems and efficiently implementing network synchronization in section 5.

## 2. Models and methods

The exploration of nonlinear systems characterized by coupled behaviors can be effectively summarized by expressing the dynamic formulation of a representative node as follows [48]:

$$\frac{dV_i}{dt} = f(V_i) + \sum_{j}^{n} w_{ij} c_{ij} g_{ij}(V_j), \tag{1}$$

where $V_i$ represents the quantity of the $i$-th node in a network of $N$ nodes that changes over time, $f(V_i)$ is the local dynamics of network nodes, $c_{ij}$ is the connectivity matrix (with a connectivity value of 1 indicating a connection and 0 otherwise), $w_{ij}$ denotes the weight of the connection, and $g_{ij}(*)$ is the coupling function. Subsequently, Eq. (1) is transformed using the Euler algorithm:

$$V_i^{t+\Delta t} = V_i^t + f(V_i^t)\Delta t + \sum_{j}^{n} w_{ij} c_{ij} g_{ij}(V_j^t)\Delta t, \tag{2}$$

where $\Delta t$ represents the time step in the numerical calculation. Subsequently, the components of the aforementioned equation, excluding the weights $w_{ij}$, are categorized as follows:

$$\begin{cases} x_{i0}^t = V_i^t + f(V_i^t)\Delta t, \\ x_{i1}^t = c_{i1} g_{i1}(V_1^t)\Delta t, \\ \quad \vdots \\ x_{in}^t = c_{in} g_{in}(V_n^t)\Delta t. \end{cases} \tag{3}$$

Arrange the formula into matrix form, denoted as $x_i^t = [x_{i1}^t, x_{i2}^t, ..., x_{in}^t]^T$, and represent the weights by $w_i^t = [w_{i1}^t, w_{i2}^t, ..., w_{in}^t]^T$. The final form of Eq. (2) is then expressed as:

$$V_i^{t+\Delta t} = w_i^T \times x_i^t + x_{i0}^t. \tag{4}$$

In the self-adaptive DLS approach, the contrast value is determined by the average of all nodes in the network that require weight adjustments at the subsequent time step:

$$\overline{V}_{t+\Delta t} = \begin{cases} \frac{1}{n}\sum_{i}^{n} V_i^{t+\Delta t}, & \text{(adaptive)} \\ V_{input}^{t+\Delta t}, & \text{(supervised)} \end{cases} \quad (5)$$

In supervised DLS, the contrast value is assigned based on the value of the external input. This set of values across all time points is represented in the form of a matrix as follows:

$$\begin{cases} \overline{V} = [\overline{V}_{t_2}, \overline{V}_{t_3}, \ldots, \overline{V}_{t_{m+1}}]^T, \\ X_i^{m \times n} = [x_i^{t_1}, x_i^{t_2}, \ldots, x_i^{t_m}]^T, \\ X_{i0} = [x_{i0}^{t_1}, x_{i0}^{t_2}, \ldots, x_{i0}^{t_m}]^T, \end{cases} \quad (6)$$

where $t_i$ represents the time point at which the variables evolve over time in equations (2) and (3). To synchronize node $i$ with the network, the objective is to minimize the square of the difference between the next calculated value from Eq. (2) and the contrast value. This minimization objective can be expressed as:

$$E = \arg\min \frac{1}{2} \sum_{t=t_1}^{t_m} (w_i^T x_i^t + x_{i0}^t - \overline{V}_{t_{m+1}})^2. \quad (7)$$

In Eq. (7), the contrast value $\overline{V}_{t_{m+1}}$, originating from either Eq. (5) or an external input and combined with $x_{i0}^t$, where $E$ represents the minimum difference between each node's value and $\overline{V}_{t_{m+1}}$ across all time points, is subsequently represented in matrix form as follows:

$$Y_i = \overline{V} - x_{i0}^t = [\overline{V}_{t_2} - x_{i0}^{t_1}, \overline{V}_{t_3} - x_{i0}^{t_2}, \ldots, \overline{V}_{t_{m+1}} - x_{i0}^{t_m}]^T. \quad (8)$$

Express Eq. (7) in matrix form, which can be represented as:

$$E = \arg\min \frac{1}{2}(X_i w_i - Y_i)^T (X_i w_i - Y_i). \quad (9)$$

Determine the gradient of the weight parameter with respect to the error, which is denoted as:

$$\frac{dE}{dw} = X_i^T (X_i w_i - Y_i). \quad (10)$$

Since the error value is obtained from the squared difference, aligning with the principles of the least squares method, the optimal value is determined by finding the minimum of this squared error. The optimal weight value is subsequently calculated as:

$$w_i = (X_i^T X_i)^{-1} X_i^T Y_i. \quad (11)$$

In practice, coupled nonlinear systems are dynamic, evolving over time and leading to an increasing series of node values. We represent this temporal evolution as follows:

$$\begin{cases} \overline{V} = [\overline{V}_{t_2}, \overline{V}_{t_3}, \ldots, \overline{V}_{t_{m+1}}]^T \to \overline{V}' = [\overline{V}_{t_2}, \overline{V}_{t_3}, \ldots, \overline{V}_{t_{m+1}}, \overline{V}_{t_{m+2}}]^T, \\ X_i = [x_i^{t_1}, x_i^{t_2}, \ldots, x_i^{t_m}]^T \to X_i' = [x_i^{t_1}, x_i^{t_2}, \ldots, x_i^{t_m}, x_i^{t_{m+1}}]^T, \\ X_{i0} = [x_{i0}^{t_1}, x_{i0}^{t_2}, \ldots, x_{i0}^{t_m}]^T \to X_{i0}' = [x_{i0}^{t_1}, x_{i0}^{t_2}, \ldots, x_{i0}^{t_m}, x_{i0}^{t_{m+1}}]^T. \end{cases} \quad (12)$$

As the system evolves, it continually generates new data, necessitating repeated computations. This process is inherently resource-intensive, making the direct application of Eq. (11) impractical for continuous use. To adapt the weight adjustment process to the system's temporal changes, Eq. (11) is reformulated into a recursive process. This enables parameter adjustment using newly acquired data. The temporal evolution of Eq. (11) is initially characterized as follows:

$$\begin{cases} R_i = X_i^T X_i \to R_i' = X_i'^T X_i', \\ Z_i = X_i^T Y_i \to Z_i' = X_i'^T Y_i', \\ w_i = R_i^{-1} Z_i \to w_i' = R_i'^{-1} Z_i'. \end{cases} \quad (13)$$

This will be followed by the application of the recursive least squares method to derive $R_i'^{-1}$ and $Z_i'$ in Eq. (13) into their respective time-varying recursive forms.

## 2.1 The recursive form of $R_i'^{-1}$

In coupled nonlinear systems with evolving time, data is continuously updated, as illustrated in Eq. (12). By employing matrix chunking calculations, the recursive relationship between matrix $R_i'$ and matrix $R_i$ can be derived as follows:

$$R_i' = X_i'^T X_i' = X_i^T X_i + x_i^{t_{m+1}}(x_i^{t_{m+1}})^T = R_i + x_i^{t_{m+1}}(x_i^{t_{m+1}})^T. \quad (14)$$

In the context of realistic coupled nonlinear systems, it is often the case that newly generated data holds greater significance than previously collected data. Consequently, upon introducing a forgetting factor ($\lambda \leq 1$) to Eq. (14), it can be formally expressed as follows:

$$R_i' = \lambda R_i + x_i^{t_{m+1}}(x_i^{t_{m+1}})^T. \quad (15)$$

Next, we solve the inverse process of Eq. (15) to derive the recursive relationship between $R_i'^{-1}$ and $R_i'$. The procedure is illustrated below:

$$R_i'^{-1} = (\lambda R_i)^{-1} - (\lambda R_i)^{-1} x_i^{t_{m+1}} (1 + (x_i^{t_{m+1}})^T (\lambda R_i)^{-1} x_i^{t_{m+1}})^{-1} (x_i^{t_{m+1}})^T (\lambda R_i)^{-1}$$

$$= \frac{1}{\lambda} R_i^{-1} - \frac{1}{\lambda} \frac{R_i^{-1} x_i^{t_{m+1}}}{\lambda + (x_i^{t_{m+1}})^T R_i^{-1} x_i^{t_{m+1}}} (x_i^{t_{m+1}})^T R_i^{-1}. \tag{16}$$

Further simplification of Eq. (16) leads to the following representation:

$$\begin{cases} P_i = R_i^{-1}, \\ P_i' = R_i'^{-1}, \\ k_i = \dfrac{R_i^{-1} x_i^{t_{m+1}}}{\lambda + (x_i^{t_{m+1}})^T R_i^{-1} x_i^{t_{m+1}}}, \\ P_i' = \dfrac{1}{\lambda} P_i - \dfrac{1}{\lambda} k_i (x_i^{t_{m+1}})^T P_i. \end{cases} \tag{17}$$

It should be further noted that the product of the two variables $P_i'$ and $x_i^{t_{m+1}}$ yields the following exact relationship:

$$P_i' x_i^{t_{m+1}} = \frac{P_i x_i^{t_{m+1}}}{\lambda + (x_i^{t_{m+1}})^T P_i x_i^{t_{m+1}}} = k_i. \tag{18}$$

After obtaining the recursive relationship for $P_i'$, $w_i$ in Eq. (13) is modified to the following expression, followed by the derivation of the recursive expressions for $Z_i'$ and $w_i'$.

$$w_i = P_i Z_i \rightarrow w_i' = P_i' Z_i'. \tag{19}$$

**2.2 The recursive form of $Z_i'$ and $w_i'$**

Similar to $R_i'^{-1}$, express the recursive relationship between $Z_i'$ and $Z_i$ as follows:

$$Z_i' = X_i'^T Y_i' = [X_i^T | x_i^{t_{m+1}}] \left[ \frac{Y_i}{Y_i^{t_{m+1}}} \right] = X_i^T Y_i + x_i^{t_{m+1}} Y_i^{t_{m+1}} = Z_i + x_i^{t_{m+1}} Y_i^{t_{m+1}}. \tag{20}$$

Similarly, introduce the forgetting factor $\lambda \leq 1$ as follows:

$$Z_i' = \lambda Z_i + x_i^{t_{m+1}} Y_i^{t_{m+1}}. \tag{21}$$

The recursive relationship between $w_i'$ and $w_i$ is derived by concatenating matrices $P_i'$ and $Z_i'$. The derivation is illustrated below:

$$\begin{aligned}
w'_i &= P'_i Z'_i \\
&= P'_i [\lambda Z_i + x_i^{t_{m+1}} Y_i^{t_{m+1}}] \\
&= \lambda [\frac{1}{\lambda} P_i - \frac{1}{\lambda} k_i (x_i^{t_{m+1}})^T P_i] Z_i + P'_i x_i^{t_{m+1}} Y_i^{t_{m+1}} \\
&= w_i - k_i [(x_i^{t_{m+1}})^T w_i - Y_i^{t_{m+1}}].
\end{aligned} \qquad (22)$$

**2.3 DLS technique**

By combining Eq. (17) and (22), the recursive DLS technique is obtained by organizing the expression as follows:

$$\begin{cases} w'_i = w_i - k_i \left[ (x_i^{t_{m+1}})^T w_i - Y_i^{t_{m+1}} \right], \\ k_i = \dfrac{P_i x_i^{t_{m+1}}}{\lambda + (x_i^{t_{m+1}})^T P_i x_i^{t_{m+1}}}, \\ P'_i = \dfrac{1}{\lambda} P_i - \dfrac{1}{\lambda} k_i (x_i^{t_{m+1}})^T P_i. \end{cases} \qquad (23)$$

Equation (23) is one of the most important results of this paper, it is an innovative DLS algorithm formulae to our knowledge. In which $x_i^{t_{m+1}} = [x_{i1}^{t_{m+1}}, x_{i2}^{t_{m+1}}, \ldots, x_{in}^{t_{m+1}}]^T$ represents the newly generated data in the nonlinear system, $w_i = [w_{i1}, w_{i2}, \ldots, w_{in}]^T$ denotes the connection weight between the $i$-th node in the nonlinear system and other nodes that require weight regulation, and $\lambda$ represents the forgetting factor. In the self-adaptive DLS, the contrast value is set to $Y_i^{t_{m+1}} = \overline{V}_{t_{m+2}} - x_{i0}^{t_{m+1}}$, where $\overline{V}_{t_{m+2}}$ represents the average of the values of all nodes that require weight adjustment at time $t_{m+2}$. Conversely, in the supervised DLS, $\overline{V}_{t_{m+2}}$ is replaced by an external input value.

The DLS technique primarily encompasses the continuous updating of learnable parameters in a nonlinear coupled system using values and contrast values generated by nonlinear oscillators, forming a recursive relationship with the previous values. As it involves a recursive process for parameter updates, an initial value is required at the initial iteration. Observing Eq. (23), the initial value $P_i$ is typically chosen as:

$$P_i = \alpha I, \qquad (24)$$

where $I$ is the identity matrix, and $\alpha$ is a hyperparameter, which is typically set to a relatively large value to prevent $P_i$ from decreasing to a negative value during the recursion.

## 3. Experiments and Related Model

In this section, two nonlinear neuron models, as well as various network structures, are described. These elements are then employed to assess the effectiveness of the DLS method proposed in this paper for learning synchronization patterns.

### 3.1 Neuronal systems

To assess the effectiveness of the proposed DLS technique, we initially employ a relatively simple coupled system consisting of nonlinear neurons known as the FitzHugh-Nagumo (FHN) model [46]. The mathematical model for the coupled FHN system is described as follows:

$$\begin{cases} \dfrac{dV_i}{dt} = V_i - \dfrac{V_i^3}{3} - W_i + I_i^{ex} + I_i^{syn}, \\ \dfrac{dW_i}{dt} = \varepsilon(V_i + a - bW_i), \\ I_i^{syn} = \sum_{j}^{n} w_{ij} c_{ij} g_{ij}(V_j), \end{cases} \quad (25)$$

where $V_i$ represents the membrane potential of the $i$-th neuron, $I_i^{ex}$ denotes the external excitation current, and $I_i^{syn}$ is the synaptic current resulting from coupling with other neurons. This coupling is characterized by connection weights $w_{ij}$, a connection matrix $c_{ij}$, and a synaptic model $g_{ij}(V_j)$. $W_i$ represents the slow recovery variable associated with sodium and potassium ion channel inactivation. Several constants in the equation are set as follows: $\varepsilon=0.08$, $a=0.7$, $b=0.8$. Eq. (25), corresponding to the DLS technique in Eq. (23), can be described as:

$$\begin{cases} x_i^{t_{m+1}} = [c_{i1} g_{i1}(V_1^{t_{m+1}})\Delta t, \ldots, c_{iN} g_{iN}(V_N^{t_{m+1}})\Delta t]^T, \\ Y_i^{t_{m+1}} = \dfrac{1}{n} \sum_{i}^{n} V_i^{t_{m+2}} - [V_i^{t_{m+1}} + (V_i^{t_{m+1}} - \dfrac{(V_i^{t_{m+1}})^3}{3} - W_i + I_i^{ex})\Delta t], \end{cases} \quad (26)$$

where $n$ represents the number of neurons for adaptive learning, and $N$ is the total number of neurons in the neural network. In this study, the time step is set to $\Delta t=0.01$, and both the forgetting factor $\lambda$, and the initial hyperparameter value $\alpha$ in the DLS technique are set to 1.

To further verify the effectiveness of the DLS technique in more complex systems, we will proceed with its verification using the Hodgkin-Huxley (HH) [47] model. The mathematical representation of the HH coupled system is as follows:

$$\begin{cases} C_m \dfrac{dV_i}{dt} = -I_{iNa} - I_{iK} - I_{iL} + I_{iext} + I_{isyn}, \\ I_{iNa} = g_{Na} m_i^3 h_i (V_i - V_{Na}), \\ I_{iK} = g_K n_i^4 (V_i - V_K), \\ I_{iL} = g_L (V_i - V_L), \\ I_{isyn} = \sum\limits_{j}^{n} w_{ij} c_{ij} g_{ij}(V_j). \end{cases} \quad (27)$$

The specific membrane capacitance of the cell membrane is denoted as $C_m$ and is set to 1 $\mu F/cm^2$. Additionally, the currents associated with sodium, potassium, and leak channels are characterized by their respective maximum conductances $g_{Na}=120mS/cm^2$, $g_K=36mS/cm^2$, and $g_L=0.3mS/cm^2$ and reversal potentials $V_{Na}=50mV$, $V_K=-77mV$, and $V_L=-54.4mV$. Furthermore, the three gating variables responsible for the behavior of sodium and potassium channels can be represented as follows:

$$\frac{dx_i}{dt} = \alpha_{x_i}(V_i)(1-x_i) - \beta_{x_i}(V_i)x_i, \quad (x=n,m,h) \quad (28)$$

The functions denoted as $\alpha_{x_i}(V_i)$ and $\beta_{x_i}(V_i)$ represent the on-off ratios of the gating variables, which are described as follows:

$$\begin{cases} \alpha_{n_i}(V_i) = \dfrac{0.01(V_i+55)}{1-\exp(-(V_i+55)/10)}, \\ \beta_{n_i}(V_i) = 0.125\exp(-(V_i+65)/80), \\ \alpha_{m_i}(V_i) = \dfrac{0.1(V_i+40)}{1-\exp(-(V_i+40)/10)}, \\ \beta_{m_i}(V_i) = 4\exp(-(V_i+65)/18), \\ \alpha_{h_i}(V_i) = 0.07\exp(-(V_i+65)/20), \\ \beta_{h_i}(V_i) = \dfrac{1.0}{1+\exp(-(V_i+35)/10)}. \end{cases} \quad (29)$$

Subsequently, Eq. (27) is transformed into the following expression according to the DLS technique:

$$\begin{cases} x_i^{t_{m+1}} = [c_{i1} g_{i1}(V_1^{t_{m+1}})\Delta t, \ldots, c_{iN} g_{iN}(V_N^{t_{m+1}})\Delta t]^T, \\ Y_i^{t_{m+1}} = \dfrac{1}{n}\sum\limits_{i}^{n} V_i^{t_{m+2}} - [V_i^{t_{m+1}} + (-I_{iNa} - I_{iK} - I_{iL} + I_{iext})\Delta t / C_m]. \end{cases} \quad (30)$$

## 3.2 Synaptic structure

In nonlinear neuronal coupling systems, the two predominant coupling modes are electrical synaptic and chemical synaptic coupling. This study initiates the verification of the DLS technique's effectiveness by employing a relatively straightforward electrosynaptic coupling method. The mathematical model of the electrical synapse in Eqs. (25) and (27) can be defined as:

$$g_{ij}(V_j) = V_j(t) - V_i(t), \tag{31}$$

where $V_j(t)$ represents the presynaptic membrane potential, and $V_i(t)$ represents the postsynaptic membrane potential. On the other hand, chemical synaptic coupling is a more intricate mode of coupling that involves processes such as presynaptic neuronal firing, chemical transmitter release, and receptor binding. For the sake of simplicity, this study adopts a widely utilized mathematical model, which can be expressed as [49]:

$$\begin{cases} g_{ij}(V_j) = \alpha(t-t_j)(V_{syn} - V_i), \\ \alpha(t) = \dfrac{t}{\tau_{syn}} \exp\left(-\dfrac{t}{\tau_{syn}}\right)\Theta(t), \end{cases} \tag{32}$$

where $t_j$ denotes the presynaptic firing time, and the synaptic reversal potential is fixed at $V_{syn}=0mV$ in this study. $\tau_{syn}$ represents the time constant, which is set to 2 ms, and $\Theta(t)$ denotes the Heaviside function.

### 3.3 Network structure

The effectiveness of the DLS technique is assessed across various network topologies in this study. Four different network topologies are employed, including fully connected (FC), small-world (SW) [20], scale-free (SF) [21], and random networks. The total number of neurons ($N$) in the study is held constant at 100.

In the FC topology, each neuron is connected to all other neurons except itself, resulting in an average degree of $N$-1. SW networks are generated using the Watts-Strogatz model [20] with an average degree of 4 and a reconnection probability of 0.4.

SF networks exhibit unique characteristics, with a few nodes having a large degree while most nodes have a small degree. These networks also have the capacity to grow. To create SF networks, the Barabási-Albert algorithm [21] is employed, initiating the network with 5 nodes.

When a new node joins the network, it connects to two existing nodes with a probability proportional to the degree of those nodes.

Finally, setting the reconnection probability to 1 in the Watts-Strogatz model results in a completely random network, where each edge is randomly reconnected to other nodes.

### 3.4 Evaluation

Monitoring the training progress is of utmost importance for assessing the performance of the DLS technique. Initially, the degree of synchronization within the network ensembles is assessed during the training phase by measuring the standard deviation:

$$e = \sqrt{\frac{1}{N}\sum_{i=1}^{N}(V_i - \frac{1}{N}\sum_{i=1}^{N}V_i)^2}. \tag{33}$$

A smaller standard deviation indicates a higher degree of collective synchronization within the network. Standard deviation provides a coarse assessment of synchronization, as its magnitude varies significantly across different systems. Therefore, it is necessary to introduce a normalized statistical measure for a more precise evaluation of synchronization. In this study, we employ a synchronization factor based on mean field theory to assess the post-training synchronization effect, which is described as follows [50]:

$$\begin{cases} F = \frac{1}{N}\sum_{i=1}^{N}V_i, \\ R = \frac{\langle F^2 \rangle - \langle F \rangle^2}{\frac{1}{N}\sum_{i=1}^{N}(\langle V_i^2 \rangle - \langle V_i \rangle^2)}, \end{cases} \tag{34}$$

where $V_i$ represents the membrane potential computed from Eqs. (25) and (27), and $\langle * \rangle$ denotes the time averaging of the variables over the computed time. The synchronization factor $R$ takes values in the range of 0 to 1, where convergence to 1 signifies synchronization and convergence to 0 signifies desynchronization.

## 4. Results

In this section, we demonstrate the learning outcomes facilitated by the dynamic learning of synchronization (DLS) technique, which includes self-adaptive and supervised mechanisms for achieving both global and local synchronization. Additionally, we explore a hybrid approach that integrates these synchronization methods. The utility of the DLS technique is further evaluated through its application to complex nonlinear neuronal models and advanced coupling strategies. For the numerical computation of membrane potentials as outlined in Equations (25) and (27), we utilize the Euler forward method with a discrete time step of 0.01. First, we use the master stability function (MSF) [51] to verify that DLS can adaptively adjust complex dynamic networks into a synchronized region.

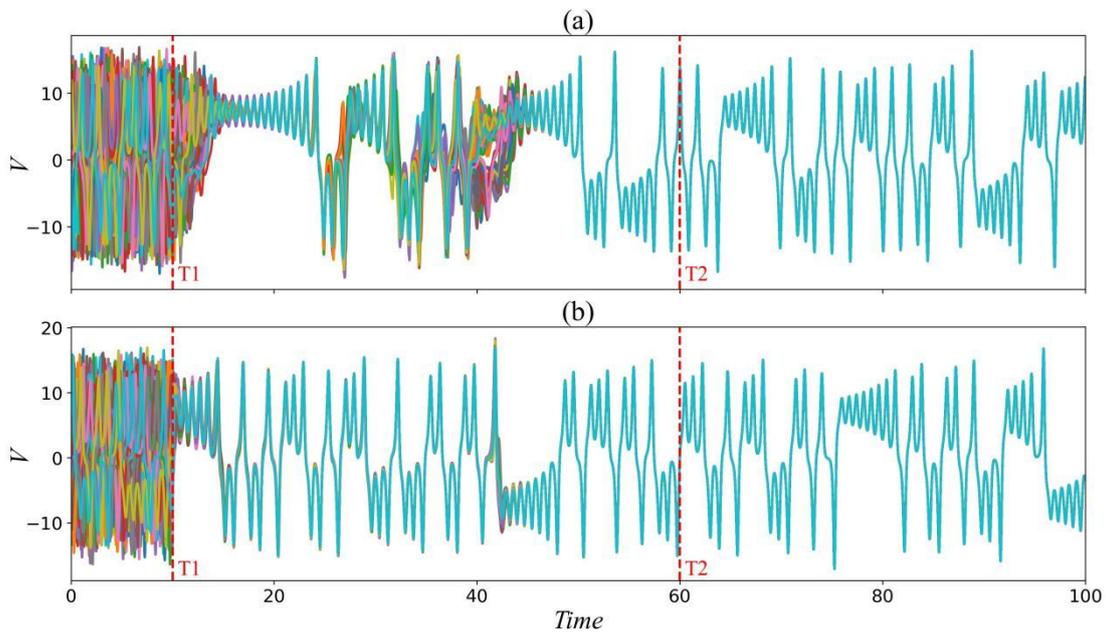

Fig. 1 The effects of the DLS in scale-free network composed of Lorenz system nodes are as follows: The nodes are governed by the Lorenz system, and the topology is scale-free. DLS is applied under two coupling schemes to regulate network synchronization, starting from T1=10 to T2=60. In (a), the coupling scheme is 1 -> 1, where the synchronization region is unbounded; in (b), the coupling scheme is 2 -> 1, where the synchronization region is bounded. The results show that in both coupling schemes, the application of DLS significantly promotes network synchronization.

**4.1 Synchronization stability analysis based on MSF**

The core concept of the MSF is to evaluate the stability of a synchronized state by analyzing the dynamics of small deviations from this state. Specifically, MSF involves examining the system's response to small perturbations and analyzing how these disturbances evolve over time.

The implementation of this method is based on a linearized analysis of the system near its synchronized state. The relevant equation (1) can be rewritten as follows:

$$\frac{dV_i}{dt} = f(V_i) - \sum_{j}^{n} l_{ij} g(V_j), \tag{35}$$

where the nodal dynamics equation is $\dot{V}_i = f(V_i)$, where $V_i \in R^n$ is the state variable for the *i*-th node. $g(V_j)$ is the internal coupling function, and it is assumed that the internal coupling relationships among the nodes are identical. The coupling matrix $L = (l_{ij})_{N \times N}$ is a weighted Laplacian matrix (meeting the dissipative coupling condition $\sum_{j} l_{ij} = 0$) and can be expressed as:

$$l_{ij} = \begin{cases} -w_{ij} c_{ij}, & i \neq j \\ \sum_{j \neq i} w_{ij} c_{ij}, & i = j \end{cases} \tag{36}$$

Let *s*(*t*) be the solution to the isolated node dynamics equation $\dot{s} = f(s)$. Denote $\xi_i(t) = x_i(t) - s(t)$ as a small perturbation to *s*(*t*). By introducing these perturbations into the system, its dynamic stability can be studied. Thus, the variational equation based on *s*(*t*) and corresponding to eq. (35) is obtained:

$$\dot{\xi}_i = Df(s)\xi_i - \sum_{j=1}^{N} l_{ij} Dg(s)\xi_j, \quad i = 1, 2, \cdots, N \tag{37}$$

where D*f*(*s*) and D*g*(*s*) respectively represent the Jacobian matrices of the functions *f*(*x*) and *g*(*x*) evaluated at the point sss. Next, let $\xi = [\xi_1, \xi_2, \cdots, \xi_N]$ and rewrite the equation accordingly. Thus, equation (37) can be rewritten as follows:

$$\dot{\xi} = Df(s)\xi - Dg(s)\xi L^T \tag{38}$$

In this setup, assume that the weighted coupling Laplacian matrix *L* is diagonalizable. Define $L^T = P\Lambda P^{-1}, \Lambda = diag(\lambda_1, \lambda_2, \cdots, \lambda_N)$, where $\lambda_k (k = 1, 2, \cdots, N)$ is the eigenvalue of the weighted coupling Laplacian matrix *L*. Setting $\eta = [\eta_1, \eta_2, \cdots, \eta_N] = \xi P$, we can describe the equation or relationship accordingly:

$$\dot{\eta}_k = [Df(s) - \lambda_k Dg(s)]\eta_k, \quad k = 2, 3, \cdots, N \tag{39}$$

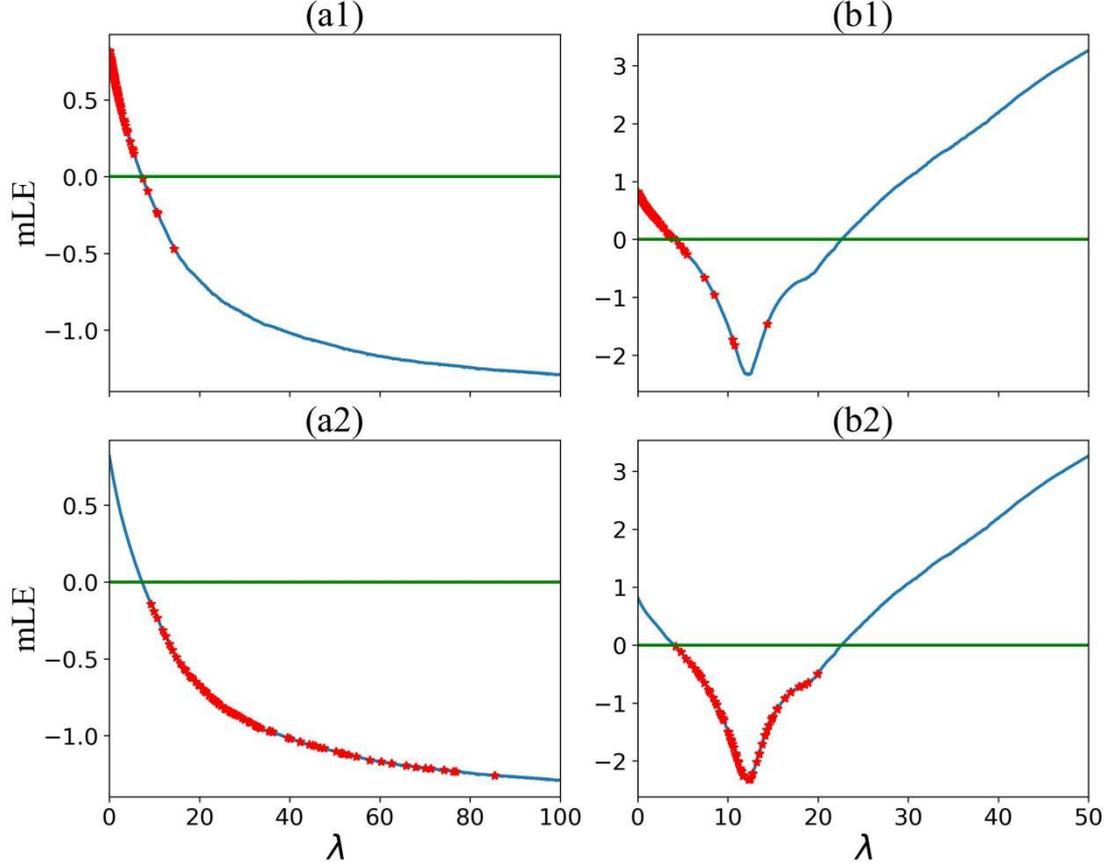

Fig. 2 The DLS effectively regulates a scale-free network composed of Lorenz systems to achieve a synchronized stable region. The red stars in the diagram represent the distribution of the eigenvalues of the weighted Laplacian matrix within the synchronization area. The (a1) and (b1) display the eigenvalue distributions under the 1->1 and 2->1 coupling schemes before DLS adjustment, respectively, while (a2) and (b2) show the distributions after DLS adjustment for the same coupling schemes. The results indicate that DLS can effectively regulate the network to a stable synchronized region.

A common criterion for assessing the stability of a synchronous manifold is to ensure that all the transverse maximum Lyapunov exponents (mLE) described in equation (39) are negative. The MSF characterizes the relationship between the mLE and the eigenvalues, and this function is defined according to equation (39). These eigenvalues are typically arranged in ascending order:

$$\varepsilon_1 < \lambda_1 < \lambda_2 \leq \cdots \leq \lambda_N < \varepsilon_2 \tag{40}$$

Next, we will apply the DLS in the Lorenz system [52] to verify that DLS can effectively adjust the network dynamics to a stable synchronization region. The Lorenz system is a classic chaotic system, and its equations can be described as follows:

$$\begin{cases} \dot{V} = \sigma(W - V), \\ \dot{W} = V(\rho - U) - W, \\ \dot{U} = VW - \beta U, \end{cases} \tag{41}$$

where the parameters are set as $\sigma = 10$, $\rho = 28$ and $\beta = 2$. For stability analysis and other calculations, the Jacobian matrix is required. The Jacobian matrix for the Lorenz system can be represented as follows:

$$Df = \begin{pmatrix} -\sigma & \sigma & 0 \\ \rho - U & -1 & -V \\ W & V & -\beta \end{pmatrix}. \tag{42}$$

In this study, we explore two different cases, specifically parameters $\varepsilon_2 < \infty$ (Type 1: bounded region) and $\varepsilon_2 < \infty$ (Type 2: unbounded region), based on the research by Huang et al [53]. These two scenarios can be realized through two types of coupling methods: 1 -> 1 and 2 -> 1. In these two cases, the Jacobian matrices D$g$ for each can be represented as follows:

$$Dg = \begin{pmatrix} 1 & 0 & 0 \\ 0 & 0 & 0 \\ 0 & 0 & 0 \end{pmatrix} or \begin{pmatrix} 0 & 1 & 0 \\ 0 & 0 & 0 \\ 0 & 0 & 0 \end{pmatrix}. \tag{43}$$

To apply the DLS in the aforementioned two cases and to verify whether DLS can effectively adjust the network to a stable synchronization region, we need to specify the input values required by DLS according to equation (41):

$$\begin{cases} x_i^{t_{m+1}} = [c_{i1} V_1^{t_{m+1}} \Delta t, \ldots, c_{iN} V_N^{t_{m+1}} \Delta t]^T & (1 \to 1), \\ x_i^{t_{m+1}} = [c_{i1} W_1^{t_{m+1}} \Delta t, \ldots, c_{iN} W_N^{t_{m+1}} \Delta t]^T & (2 \to 1), \\ Y_i^{t_{m+1}} = \frac{1}{n} \sum_i^n V_i^{t_{m+2}} - [V_i^{t_{m+1}} + (\sigma(W_i^{t_{m+1}} - V_i^{t_{m+1}}))\Delta t], \end{cases} \tag{44}$$

For the numerical simulation of the Lorenz system, we use a fourth-order Runge-Kutta method with a step size of 0.01. The initial values for each variable, $V$, $W$, and $U$, are randomly set between 0 and 1. The entire simulation encompasses a network consisting of 100 nodes, which are interconnected based on the aforementioned scale-free topological structure. Additionally, the initial weights are also randomly set between 0 and 1, to simulate the network's dynamic behavior effectively.

Firstly, this study investigates whether the application of the DLS can maintain synchronized dynamics in networks of Lorenz nodes using two different coupling schemes. As shown in Fig. 1, the application period of DLS spans from T1=10 to T2=60. Before activating DLS, the network clearly displayed asynchronous dynamics. However, after the application of DLS, the synchronization of the network was stably maintained. According to the theory of the MSF, all eigenvalues of all network structures must fall within the synchronization region, which is a necessary condition for achieving local complete synchronization. In other words, synchronization can only proceed stably when the eigenvalues are within the synchronization region. Furthermore, to verify whether DLS adjustment can stabilize network synchronization, this study explore the distribution of the eigenvalues of the weighted Laplacian matrix in the synchronization region before and after the application of DLS, with related results displayed in Fig. 2.

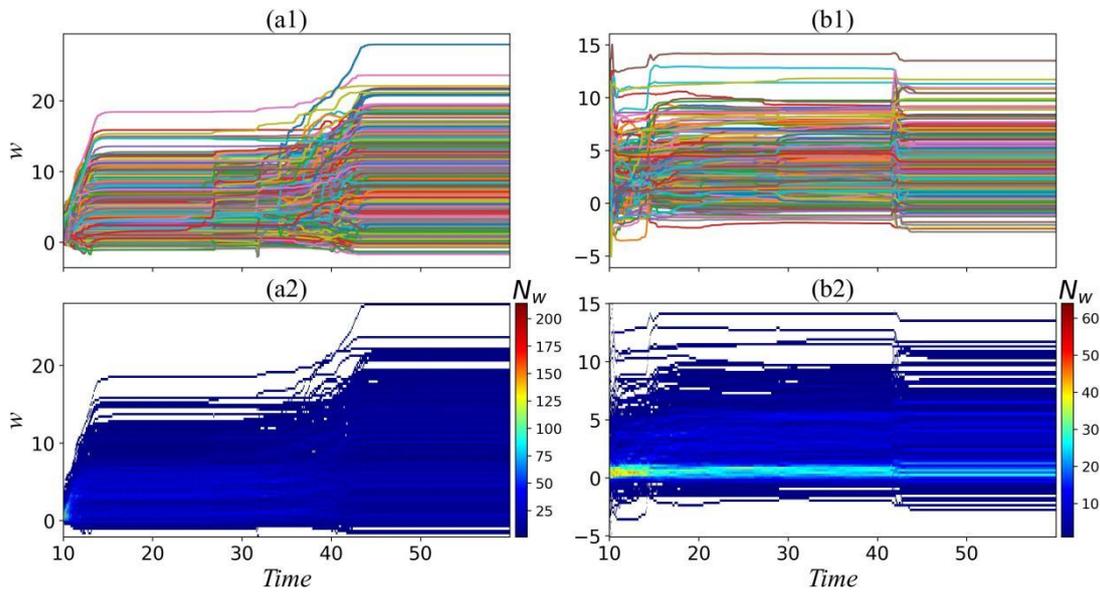

Fig. 3 The DLS adjusts the changes in connection weights within a scale-free network composed of Lorenz systems. (a) and (b) respectively illustrate the changes in weights during DLS adjustment under the 1->1 and 2->1 coupling schemes. (a1) and (b1) display the variation of all connection weights over time, while (a2) and (b2) show the time-series histograms of the weights, which depict the changes in weight distribution over time. The results indicate that after a period of DLS adjustment, the connection weights tend to stabilize and show reduced variability.

Fig. 2 displays the initial results, with Figs. 2(a1) and (b1) illustrating the unbounded and bounded cases of the synchronization region under two different coupling schemes. The red stars in the figures mark the distribution of the weighted Laplacian matrix weights at the initial time. Figs. 2(a2) and (b2) show the results after DLS adjustment, demonstrating that all eigenvalues have been effectively adjusted within the synchronization region. Combined with the results from Fig. 1, this indicates that DLS can effectively regulate network synchronization and maintain stable synchronization of the network. Next, this paper discuss the changes in weights during the DLS adjustment process, with related results displayed in Fig. 3.

Fig.s 3(a1) and (b1) show the changes in each weight *w* over time during the DLS adjustment process under two different coupling schemes. Figs. 3(a2) and (b2) display the changes in weight distribution over time through time series histograms. The results indicate that during the DLS adjustment, the weights enter an optimal parameter space and eventually converge to a stable value, after which no significant changes occur. Notably, the results in Fig. 3 show that DLS adjusts some weights to become negative. Although negative weights are generally considered detrimental to synchronization, the results in Fig. 2 demonstrate that DLS can adjust the eigenvalues of the weighted Laplacian matrix to the synchronization region. This suggests that even in the presence of negative weights, DLS can still effectively adjust the weights to the optimal parameter space without affecting the overall synchronization effect.

In the aforementioned study, we explored the effects of using the DLS to regulate network synchronization in a scale-free network composed of Lorenz systems, employing two different coupling schemes. The MSF theory indicates that DLS can effectively adjust the network into a synchronization region, thereby achieving stable synchronous dynamics. Building on these results, this study next apply DLS to networks composed of nonlinear neuron models, aiming to evaluate the significant impact of DLS on network synchronization regulation.

**4.2 Global synchronization**

The DLS technique's effectiveness for achieving self-adaptive global synchronization within the network is first verified. Within the context of adaptive synchronization, $\overline{V}_{t_{m+2}}$ of the contrast value $Y_i^{t_{m+1}} = \overline{V}_{t_{m+2}} - x_{i_0}^{t_{m+1}}$ in Eq. (23) signifies the average membrane potential of all neurons in the network undergoing synchronization learning. Fig. 4 displays the validation outcomes of the DLS effect, employing the FitzHugh-Nagumo (FHN) model, a fully connected (FC) network structure, and electrical synaptic coupling. For all FHN models in the study, the initial values of the variables in Eq. (24) are set to random numbers between 0 and 1. The external excitation currents $I_i^{ex}$ are Gaussian-distributed with a mean and standard deviation of 1, whereas the connection weights $w_{ij}$ are uniformly distributed random numbers between -0.2 and 0.2. The experimental methodology, depicted in Fig. 4, unfolds in six phases: (1) initialization, (2) training, (3) testing, (4) disruption (attacking), (5) retraining, and (6) retesting, with transformation times set at T1=300, T2=600, T3=1100, T4=1600, and T5=1900. During the disruption phase, any connections with weight values exceeding 0.15 are severed, thereby inducing desynchronization within the network once again.

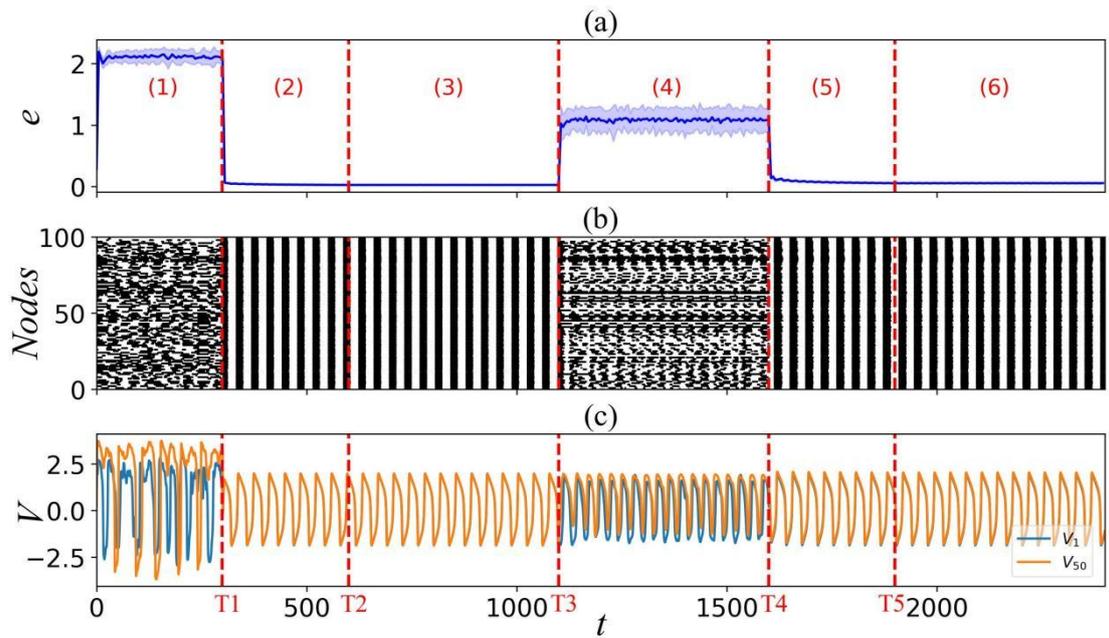

Fig. 4 Effectiveness of the DLS technique in FC networks. The experiment, utilizing the FHN model, FC network structure, and electrical synaptic coupling, is divided into six stages: (1) initialization, (2) training, (3) testing, followed by (4) attack--where connections exceeding a weight of 0.15 are disconnected to induce desynchronization, (5) retraining, and (6) retesting. These stages are repeated 20 times. Key time points T1, T2, T3, T4, and T5 are set at 300, 600, 1100, 1600, and 1900, respectively. (a) plots the standard deviation changes in membrane potential over the 20 repetitions. (b) depicts the firing raster diagram from one of these repetitions. Additionally, (c) demonstrates the fluctuations in membrane potential of the network's 1st and 50th neurons. Notably, during the two testing phases, the average synchronization coefficients, obtained from 20 repetitions, are measured at 0.99 ($\pm$2.04e-08) and 0.99 ($\pm$1.50e-06), respectively, demonstrating the consistent and effective synchronization capability of the DLS technique.

The results, derived from averaging the results of 20 trials for each condition, conclusively demonstrate the DLS technique's efficacy. In Fig. 4(a), the application of the DLS technique is shown to significantly and swiftly reduce the standard deviation of the membrane potential, leading to its stabilization. Moreover, when the network is disrupted, the DLS method proficiently restores synchronization. Figs. 4(b) and (c) visually depict this effect in one of the trials, unequivocally proving the neural network's synchronization achievement. To further verify the network's attainment of a synchronized state post-DLS application, synchronization factors for both test phases were computed. The computed average synchronization factors, 0.99 (±2.04e-08) and 0.99 (±1.50e-06) across 20 trials for the respective phases, reinforce the DLS technique's success. Additional experiments in complex network structures, such as small-world, scale-free, and randomized networks demonstrated in Fig. 5, further attest to the DLS technique's robustness.

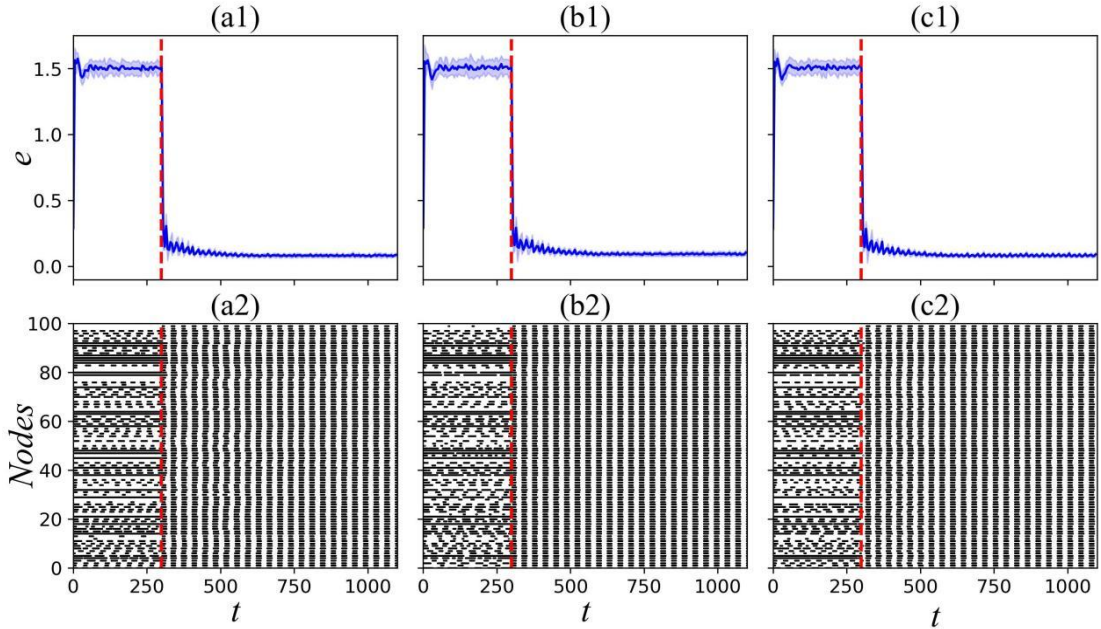

Fig. 5 Effectiveness of DLS technique in complex networks. This figure illustrates the outcomes of the DLS technique when applied to small-world, scale-free, and randomized networks, depicted in (a), (b), and (c), respectively. The DLS technique is activated from a start time of 300 and concludes at 600, with this procedure being repeated 20 times. The standard deviations for each network type, compiled from these 20 repetitions, are shown in (a1), (b1), and (c1). Additionally, visualizations from one of these repetitions for each network type are presented in (a2), (b2), and (c2). The consistent effectiveness of the DLS technique in these complex network scenarios is thereby demonstrated.

The DLS technique's efficacy in complex networks is validated through averaging results from conducting the procedure 20 times for each setting, as illustrated in Fig. 5. In Fig. 5 (a1), (b1), and (c1), a consistent decrease in the standard deviation of the neuron potentials across the network is observed following the DLS technique's application. The establishment and exceptional steadiness of the synchronization pattern are clearly showcased in the visual displays of Fig. 5 (a2), (b2), and (c2). Further exploration into the impact of the DLS technique on weight distribution across various networks is conducted, with the findings represented as a histogram in Fig. 6.

The findings presented in Fig. 6 elucidate the transformation in weight distribution following the application of the DLS technique. Initially characterized by uniform distribution, the weights, post-DLS application, display a pronounced central concentration, tapering off symmetrically towards the ends -- a pattern reminiscent of a Gaussian curve. This shift underscores the DLS technique's role in facilitating network synchronization. Specifically, in FC networks, the distribution not only becomes more uniform but also shifts towards lower values relative to other network models, a change attributable to the dense connectivity inherent to FC networks. Upon comparing small-world networks (illustrated in Fig. 6b) with scale-free networks (Fig. 6c), it becomes apparent that scale-free networks feature a significantly larger fraction of weights in the negative spectrum. This tendency can be linked to the scale-free network's densely interconnected nodes, necessitating a greater prevalence of negative weights to maintain balance. Further analysis into the DLS technique's performance in expansive networks is documented in Fig. 7, offering insights into its scalability and efficacy across various network sizes.

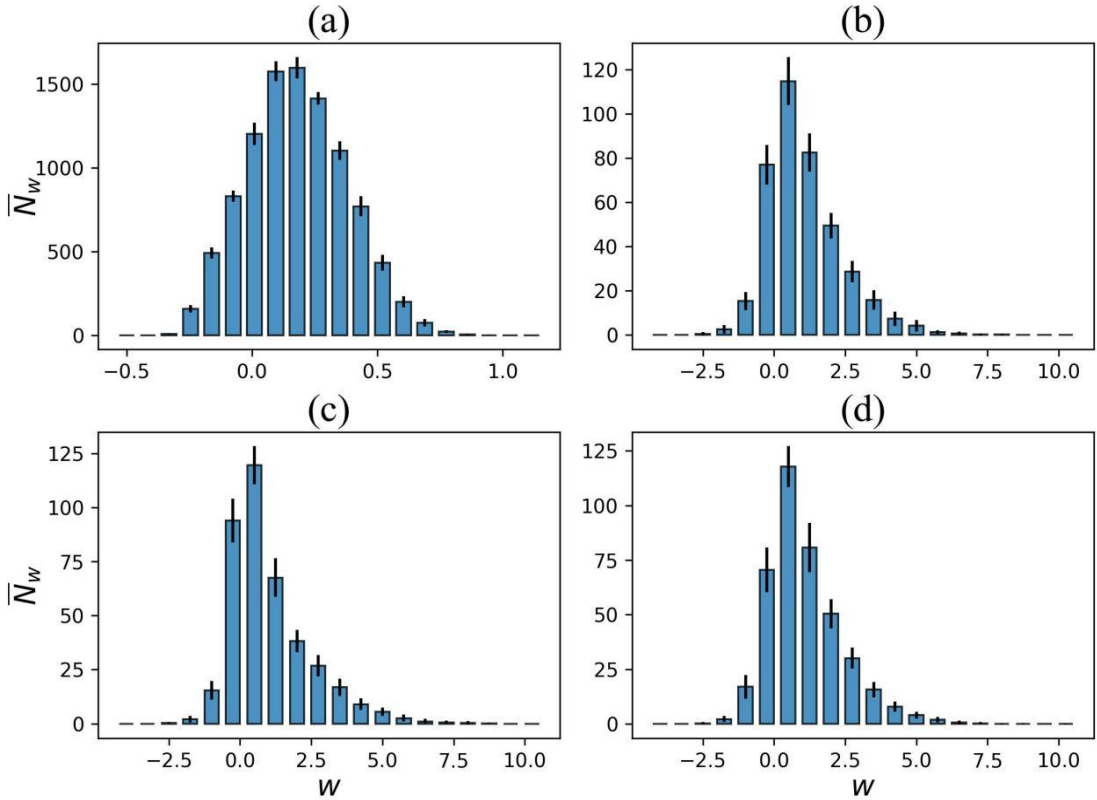

Fig. 6 Histogram of weight distribution in different networks following DLS technique application. The figure illustrates the weight distributions in Fully Connected (FC), small-world, scale-free, and randomized networks after employing the DLS technique, as depicted in (a), (b), (c), and (d), respectively. Following the application of the DLS technique, the weight distribution in each network type becomes more concentrated around a central peak and diminishes towards both ends, resembling a Gaussian-like curve.

Fig. 7(a) showcases the synchronization performance across networks of various sizes, as determined by the synchronization factor detailed in Eq. (33), following the deployment of the DLS technique. The findings uniformly underscore DLS's substantial beneficial impact on complex networks, irrespective of their size. A notable observation is the synchronization factors nearing 1 in the graphical representations, signaling an optimally synchronized state. Fig. 7(b) showcases the changes in the standard deviation of neuronal membrane potentials within a 500-node network. Here, the DLS application, spanning from 300 ms to 600 ms, yields comparably favorable results in networks of larger sizes, affirming DLS's robustness. Fig. 7(c) displays a firing raster diagram for a scale-free network, treated with DLS, comprising 500 nodes. This visualization effectively captures the network's shift from desynchronization towards synchronization, offering a vivid depiction of DLS's capability to harmonize network activity.

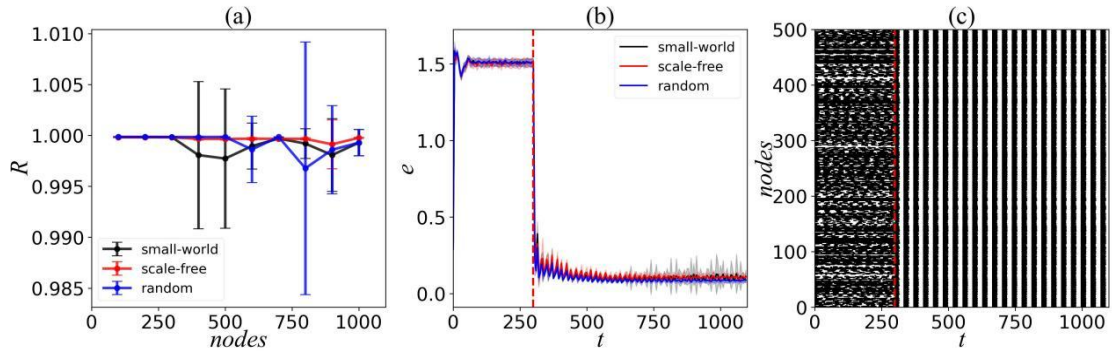

Fig. 7 Effectiveness of DLS in complex networks across various scales. This figure demonstrates the DLS technique's impact on complex networks of different sizes. (a) shows the synchronization factors for these networks following the application of DLS. In (b), the changes in the standard deviation of neuronal membrane potential due to DLS action in a network scaled to 500 nodes are plotted. Additionally, (c) presents the firing raster diagram for a single application of DLS in a scale-free network with 500 nodes. These results collectively indicate that DLS exerts a significant influence on complex networks, regardless of their scale.

## 4.3 Local synchronization

Investigating local synchronization within a network emerges as a pivotal area of study. This aspect is explored through an experiment assessing the DLS technique's capability to induce local synchronization within the network. The FC network model is selected for this purpose, with DLS being applied specifically to nodes 1 through 60 or 90. For measuring local synchronization, besides measuring the standard deviation between nodes within the synchronization region, we also separately measure the synchronization factor for areas where DLS has been applied and areas without DLS application. Equation (34) can be modified to:

$$\begin{cases} F = \dfrac{1}{N}\sum_{i=N_1}^{N_2} V_i, \\ R_{N_1-N_2} = \dfrac{\langle F^2 \rangle - \langle F \rangle^2}{\dfrac{1}{N}\sum_{i=N_1}^{N_2}(\langle V_i^2 \rangle - \langle V_i \rangle^2)}, \end{cases} \quad (45)$$

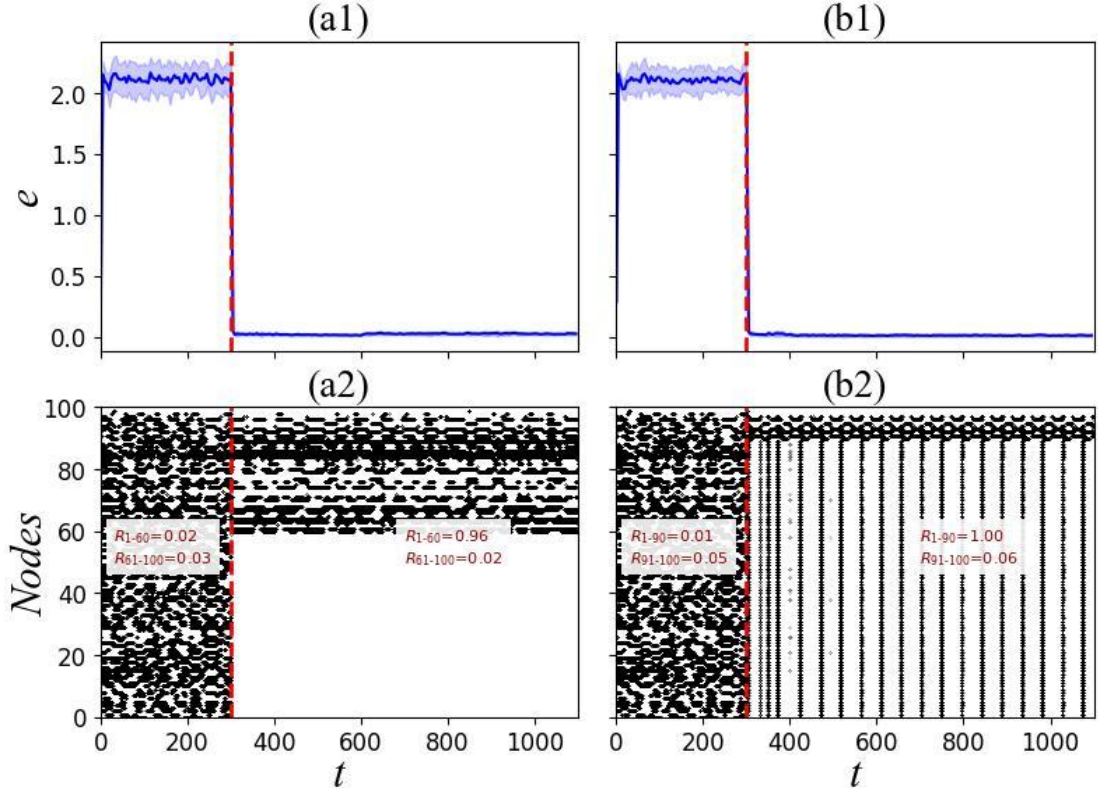

Fig. 8 Local synchronization effects of adaptive DLS in FC networks. The figure highlights the results of applying adaptive DLS to specific nodes in Fully Connected (FC) networks. The impact of DLS on nodes 1 through 60 and on node 90 is depicted in (a) and (b), respectively. A key aspect of this figure is (a1) and (b1), which prominently display the changes in standard deviation of neurons within the DLS application areas, effectively showcasing the degree of local synchronization achieved. Additionally, (a2) and (b2) provide representative firing raster diagrams, further illustrating the synchronization effects. These results underscore the ability of adaptive DLS to attain significant local synchronization in targeted node ranges within FC networks.

In Equation (45), the synchronization factor is measured for local nodes $N_1$ to $N_2$. The outcomes of this application are visually represented in Fig. 8. Fig. 8(a1) and (b1) depict the variations in neuronal standard deviation over time in areas targeted by DLS, revealing a swift reduction in standard deviations after DLS is introduced. Furthermore, the firing raster diagrams in Fig. 8(a2) and (b2) illustrate that applying DLS to nodes 1 to 60 leads to an absence of firing within that segment, generating a pattern akin to a chimeric state. In contrast, applying DLS across nodes 1 to 90 showcases a marked local synchronization effect. To offer a detailed view on the impact of DLS on local synchronization, Fig. 9 compiles data on synchronization factors and firing rates -- the latter measured as the average number of spikes per time unit -- following the use of DLS. This comprehensive analysis highlights the nuanced effects of DLS on enhancing network synchronization, emphasizing its efficacy at both localized and network-wide scales.

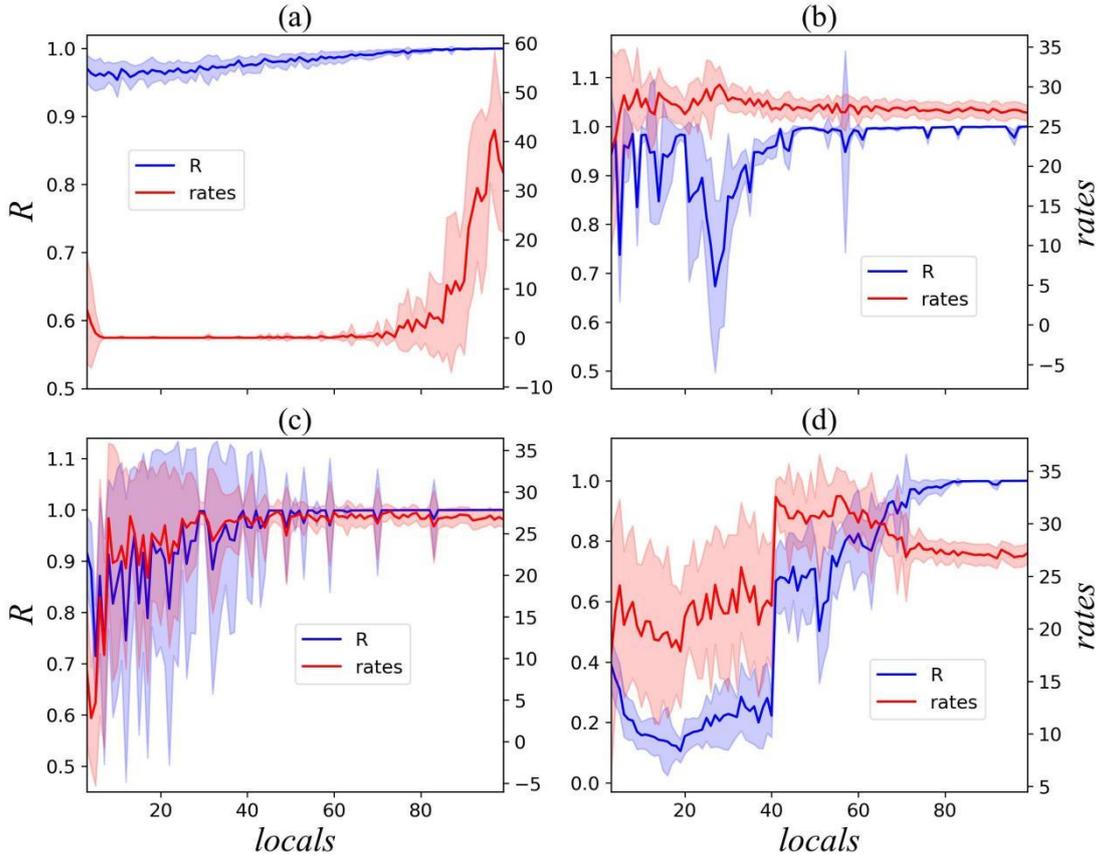

Fig. 9 DLS application in various regions of complex networks. This figure demonstrates the effects of applying DLS in diverse regions within complex network types. (a), (b), (c), and (d) respectively show the synchronization factor ($R$) and firing rates (*rates*) in the DLS application regions during the test phase after local learning in FC, small-world, scale-free, and random networks. The horizontal axis in each panel indicates the scale of the DLS application area, from 0 to a large extent. The results clearly indicate that a significant local synchronization effect is achieved in all cases where the DLS application area is large.

Acknowledging instances where neuronal activity may cease despite elevated synchronization factors, we have adopted the firing rate as a pivotal metric for the meticulous evaluation of local synchronization phenomena. Fig. 9 details the influence of the DLS technique on local regions within various network configurations, including FC, small-world, scale-free, and random networks. The findings in Fig. 9(a), consistent with those observed in Fig. 8, reveal that in FC networks, applying DLS to a restricted area results in the inactivity of neurons. Nonetheless, the application of DLS over more extensive regions within these networks culminates in successful local synchronization. Conversely, networks characterized by sparser connectivity patterns display a diminished propensity for inactivity in specific regions. As demonstrated in Figs. 9(b) and (c), small-world and scale-free networks exhibit enhanced feasibility for achieving local synchronization with DLS applied to narrower regions, albeit potentially compromising the overall synchronization quality within the DLS-affected zone. Furthermore, Fig. 9(d) reveals that in random networks, restricted application of DLS consistently precipitates suboptimal local synchronization outcomes.

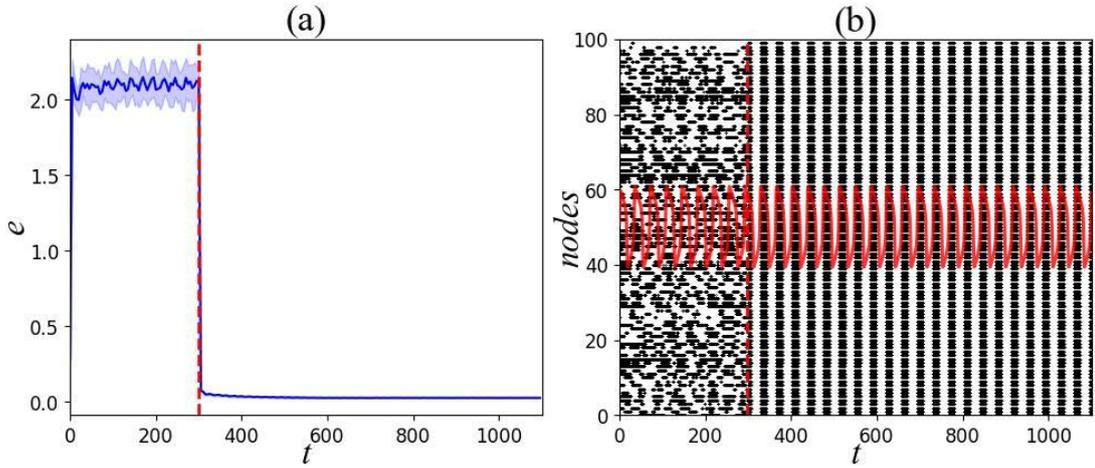

Fig. 10 Effectiveness of global supervised DLS technique in FC networks. (a) displays the averaged changes in the standard deviation of membrane potentials, calculated from 20 repeated trials, demonstrating the impact of the DLS technique in FC networks. (b) presents a firing raster diagram coupled with the supervised values, marked by a red solid line. The application of the DLS technique enables the network's neurons to synchronize their firing frequencies with these supervised values, highlighting the technique's effectiveness in achieving synchronization.

### 4.4 Supervised DLS

In the context of our analysis of adaptive DLS, Eq. (23) assumes a pivotal role where $\overline{V}_{t_{m+2}}$ in $Y_i^{t_{m+1}} = \overline{V}_{t_{m+2}} - x_{i0}^{t_{m+1}}$ represents the average value of all nodes in the network at a given moment. Subsequent to the attainment of synchronicity within the network, the modulation of its firing rate presents a considerable challenge. To mitigate this limitation, we devised a variant methodology, designated as supervised DLS. Within this methodology, the variable $\overline{V}_{t_{m+2}}$ is substituted by a parameter that is subject to external supervision. This methodological entails the substitution of with an externally prescribed, supervised value. During the application of supervised DLS, equation (26) can be modified to:

$$\begin{cases} x_i^{t_{m+1}} = [c_{i1}g_{i1}(V_1^{t_{m+1}})\Delta t, \ldots, c_{iN}g_{iN}(V_N^{t_{m+1}})\Delta t]^T, \\ Y_i^{t_{m+1}} = V_{input}^{t_{m+2}} - [V_i^{t_{m+1}} + (V_i^{t_{m+1}} - \frac{(V_i^{t_{m+1}})^3}{3} - W_i + I_i^{ex})\Delta t]. \end{cases} \quad (46)$$

$V_{input}^{t_{m+2}}$ represents the value of external input, specifying the membrane potential of neurons that require a specific firing frequency. To ensure experimental consistency and authenticity, identical initial conditions are maintained across both the adaptive and supervised modalities within the DLS framework. The distinction arises in the context of applying supervised DLS to networks constituted by the FHN model, where the external supervision value is designated as the membrane potential of a specific single node, with this node's external stimulation fixed at 1. Fig. 10 demonstrates the efficacy of the DLS technique within a FC network, evidencing its capability to facilitate network synchronization effectively.

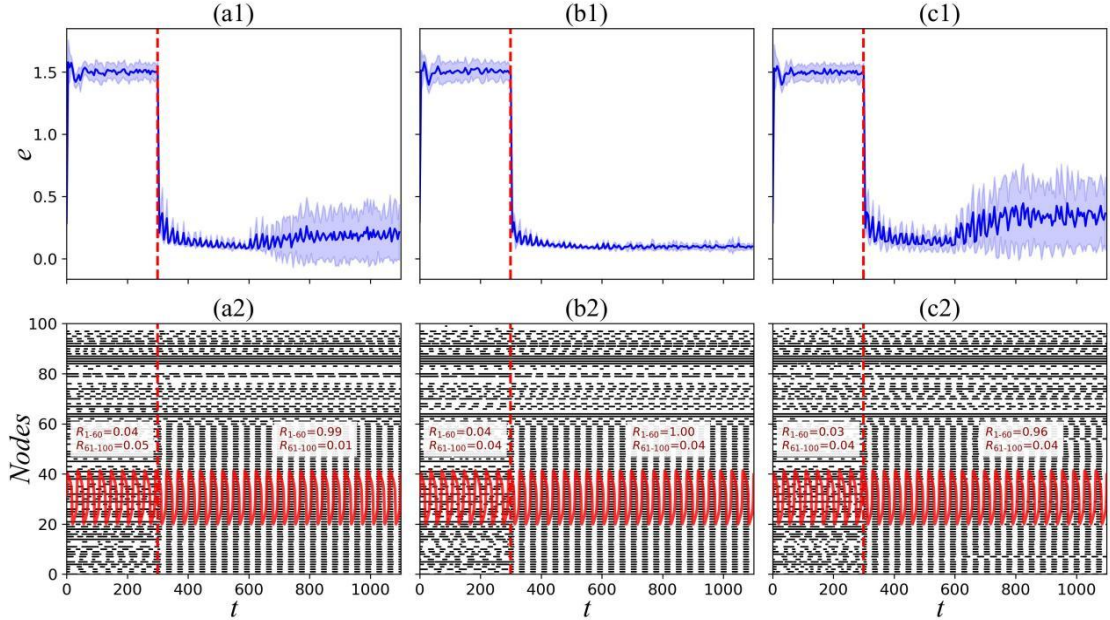

Fig. 11 Effectiveness of locally supervised DLS technique in complex networks. (a), (b), and (c) display the outcomes of applying supervised DLS to nodes 1 to 60 in small-world, scale-free, and random networks, respectively. The changes in both average values and standard deviations across these networks due to DLS application are shown in (a1), (b1), and (c1). Visual representations in (a2), (b2), and (c2), with red solid lines denoting supervised values, are also provided. The data clearly demonstrate the effectiveness of localized supervised DLS in achieving local synchronization within these complex network structures.

The data depicted in Fig. 10 underscore the pronounced efficacy of the supervised DLS approach. Notably, Fig. 10(a) showcases a swift reduction in the network's standard deviation, paralleling the results attained through the adaptive DLS method. The visual data in Fig. 10(b) corroborate that the final synchronization state corresponds with the firing frequency dictated by the externally supervised values. To further elucidate the influence of supervised DLS, examinations of its applicability and performance in targeted regions within complex network structures were undertaken. The results of these investigations are succinctly illustrated in Fig. 11, affirming the method's effectiveness in fostering local synchronization.

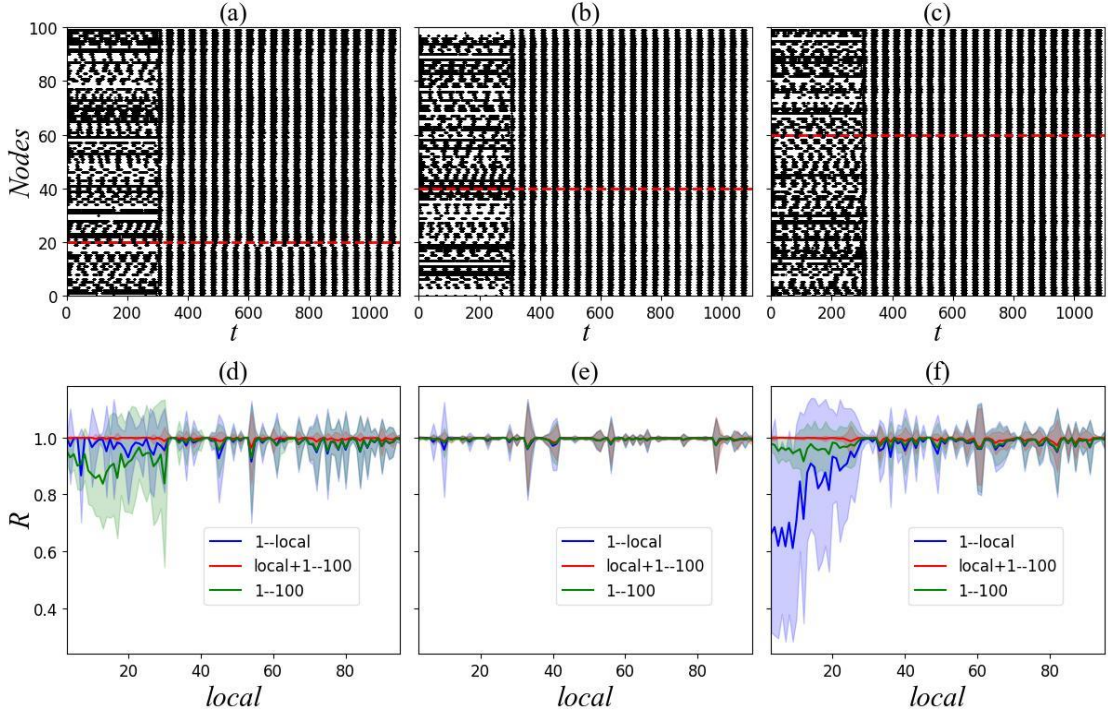

Fig. 12 Hybrid action of supervised and self-adaptive DLS in complex networks. Supervised DLS targets nodes 1 to a specific *local* node, while self-adaptive DLS extends from *Local*+1 to node 100. (a), (b), and (c) illustrate the effects in small-world networks at locals 20, 40, and 60. (d), (e), and (f) depict synchronization factor changes in small-world, scale-free, and random networks. These results showcase the synergistic impact of combining supervised and self-adaptive DLS approaches in different network contexts.

Fig. 11 displays the results of applying supervised DLS to nodes 1 through 60 across small-world, scale-free, and random networks. Figs. 11(a1) and 8(c1) show a slight reduction in synchronization performance in small-world and stochastic networks, which is also evident from the visual results in Figs. 11(a2) and 8(c2). In contrast, the scale-free network exhibits stable performance, as indicated in Figs. 11(b1) and 8(b2). This could be attributed to the structure of scale-free networks, where a few nodes hold the majority of the weights, thus facilitating easier synchronization. Moreover, the application of locally supervised DLS in conjunction with adaptive DLS in a hybrid training setup shows promising results. This approach, where adaptive

DLS supplements regions not addressed by locally supervised DLS, is effectively demonstrated in Fig. 12.

The efficacy of the hybrid DLS methodology is firstly evaluated within a small-world network framework. Figs. 12(a), (b), and (c) provide graphical representations of the supervised DLS implementation targeting nodes 1 to 20, 40, and 60, respectively. Observations from Fig. 12(a) reveal that, while distinct ranges are capable of achieving synchronized states, these states do not coincide across the different segments. Conversely, Figs. 12(b) and (c) illustrate notable synchronization across the entire network, highlighting the hybrid DLS technique's comprehensive impact. To delve deeper into the dynamics of hybrid DLS, its deployment across varied node ranges within small-world, scale-free, and stochastic networks is depicted through Figs. 12(d), (e), and (f), offering a detailed examination of its adaptability and effectiveness across diverse network topologies.

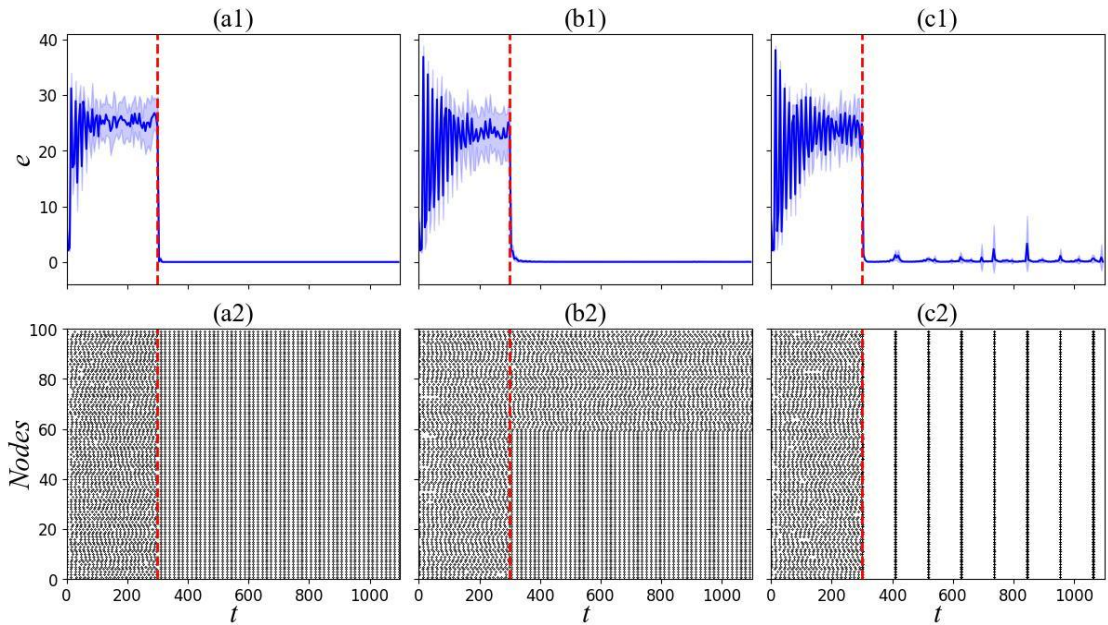

Fig. 13 Effectiveness of DLS in HH neuronal coupled systems. This figure demonstrates the impact of applying DLS in Hodgkin-Huxley (HH) networks with different synaptic connections. In (a1) and (a2), the influence of DLS on an HH network with electrical synapses is shown through the standard deviation between neurons and the corresponding firing raster diagrams. (b1) and (b2) depict the effects of adaptive DLS on local synchronization in scale-free networks. For networks connected by chemical synapses, the results are presented in (c1) and (c2). Across these varied configurations, DLS consistently achieves significant outcomes in complex coupled systems.

Fig. 12(d) shows that in smaller ranges of supervised DLS application, both supervised DLS and self-adaptive DLS synchronize better, while the network as a whole is less synchronized, demonstrating the desynchronization effect between the two ranges. The network achieves stable synchronization as the range of supervised DLS increases. Fig. 12(e) reveals that in scale-free networks, the whole network remains steadily synchronized. Fig. 12(f) indicates that in random networks, where supervised DLS is applied to a smaller extent, the synchronization of the adaptive application range maintains good synchronization, while supervised DLS is less effective. Across the three complex network types, as the range of supervised DLS applications becomes progressively larger, significant synchronization patterns emerge.

## 4.5 Synchronization of complex systems

To extend the verification of the DLS technique's efficacy, we engage with a more complex nonlinear system and coupling strategy, as specified in Equations (27) and (32). Within the scope of the Hodgkin-Huxley (HH) model analysis, the initial conditions for the variables $V$, $m$, $n$, and $h$ are uniformly and randomly assigned within ranges of -0.3 to 0.3, 0 to 0.5, 0 to 1, and 0.6, respectively. The weight parameter $w$ is initialized within the range -0.02 and 0.02. For electrical synapses, the external excitation is modeled to follow a Gaussian distribution, characterized by a mean of 10 and a standard deviation of 0.8, while for chemical synapses, the standard deviation is adjusted to 0.08 The definitive outcomes of this comprehensive investigation are depicted in Figs. 13 and 14.

Fig. 13 includes the application of the DLS technique for achieving both global and local synchronization in HH neuronal networks, connected either by electrical or chemical synapses. The data in Fig. 13(a1) and (a2) demonstrate that significant synchronization states are attainable using self-adaptive DLS in a FC network of HH neurons with electrical synapses. Fig. 13(b1) and (b2) depict the local synchronization effects within a scale-free network, highlighting that the standard deviation of the membrane potential in the DLS application area decreases rapidly. Additionally, the firing raster diagram clearly visualizes a substantial local synchronization effect. The impact of applying DLS to a chemosynaptically coupled FC network is presented in Figs. 13(c1) and (c2), where a significant synchronization effect is observed, albeit with a considerable reduction in the frequency of neuronal firing. Lastly, the efficacy of the DLS technique, when applied to HH neurons connected by either electrical or chemical synapses, is confirmed in small-world, scale-free, and random networks.

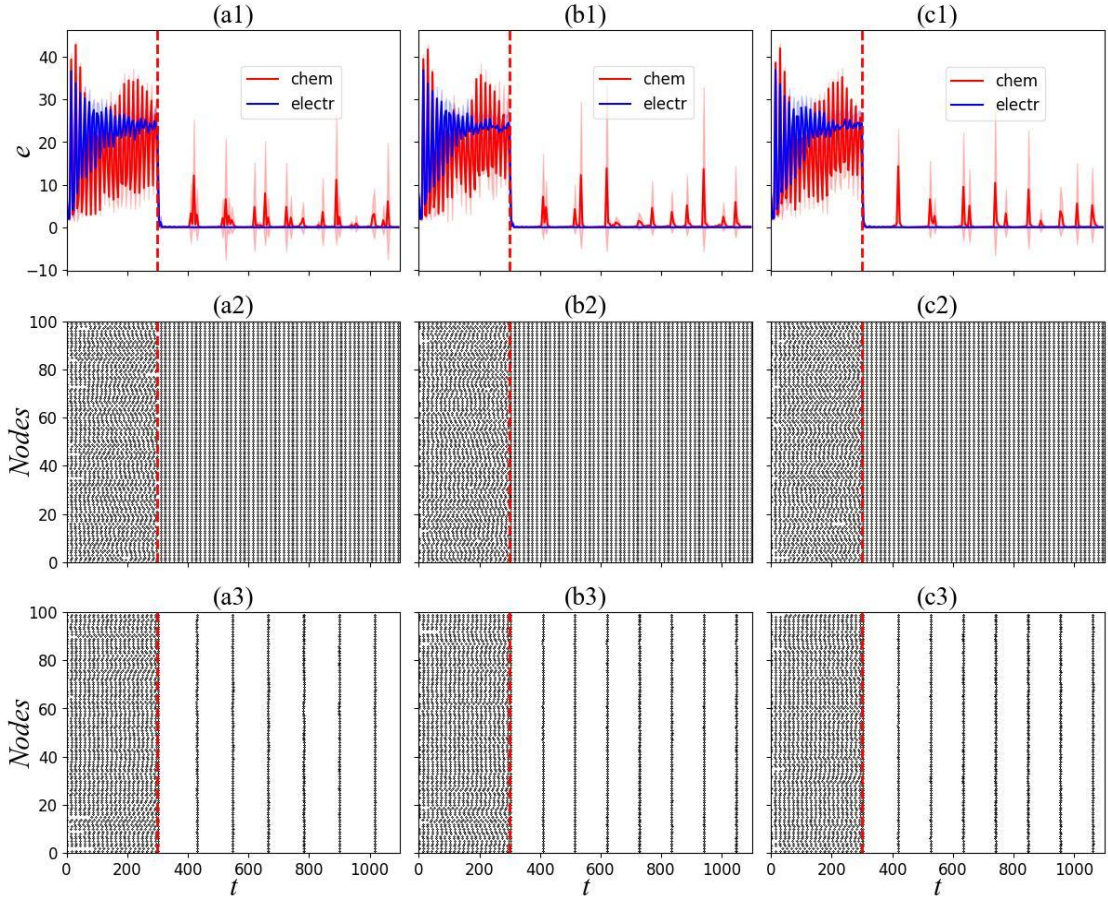

Fig. 14 Effectiveness of DLS in HH neural networks with electrical or chemical couplings. This figure highlights the outcomes of applying DLS to Hodgkin-Huxley (HH) neural networks, each with distinct complex network structures. Panels (a), (b), and (c) display the DLS results in small-world, scale-free, and stochastic networks, respectively. The time-evolving standard deviation of neuronal activity is shown in subpanels (1), while the firing raster diagrams for networks with electrical and chemical synaptic connections are depicted in subpanels (2) and (3), correspondingly. These findings illustrate that DLS is capable of achieving notable performance across various complex network systems.

The findings presented in Fig. 14 indicate that the DLS technique is effective in complex networks, utilizing sophisticated nonlinear dynamic models and intricate coupling methods. The variation in standard deviation of membrane potential, as shown in Fig. 14, reveals that networks of HH neurons coupled by chemical synapses experience fluctuations, while those connected by electrical synapses exhibit remarkable stability. The analysis of the firing raster diagrams in these complex networks reveals that after the application of DLS, neuronal networks coupled via electrical synapses retain high firing rates. Conversely, those coupled by chemical synapses show a decrease in firing rates. This collective evidence underscores the DLS technique's ability to foster significant synchronization within the network.

## 4.6 Discussion and outlook

In this article, we have developed a technique to promote synchronization in complex networks, named Dynamic Learning for Synchronization (DLS). This method captures the differences in state between nodes within the network and effectively adjusts network weights to achieve node synchronization. Utilizing the Master Stability Function (MSF), we have demonstrated that DLS can effectively adjust the network into a synchronized region, ensuring the stability of the synchronization process. We tested DLS in neural networks, and the results show that DLS can effectively regulate network synchronization.

DLS is based on the recursive least squares method, a discretization approach. Before applying this method, a learning rate must be set, a hyperparameter that significantly influences the adjustment results. Future research should consider exploring the use of adaptive learning rates to replace fixed rates. Additionally, to improve computational efficiency, the Euler method was used in the discretization of nonlinear coupled systems, which compromised computational accuracy. In subsequent research, it would be advisable to use higher precision algorithms for improvement.

DLS adjusts weights by capturing the state variables between nodes, ultimately achieving stable network synchronization. Given the wide applicability of the recursive least squares method, DLS can be used in most models, for example, to adjust synchronization in higher-order primitive networks [54]. Future studies should consider using DLS in different complex networks to find the optimal parameter space for synchronization adjustment.

DLS is a real-time method for adjusting network parameters, capable not only of adjusting network weights but also of dynamically regulating external factors input into complex networks, such as external stimuli in neuronal networks [55], thus achieving continuous correction of network synchronization. Through this mode of adjustment, the application of DLS in future research will become more widespread, for instance, adaptively adjusting external factors to eliminate certain adverse collective network behaviors, such as spiral waves.

## 5. Conclusions

Inspired by the weight regulation method in artificial neural networks (ANN), this study introduces a dynamical learning of synchronization (DLS) technique designed to control the synchronization state of a complex network composed of nonlinear oscillators. The DLS technique comes in two forms: self-adaptive DLS, where the network autonomously adjusts its synchronization, and supervised DLS, where external inputs compel the network to achieve a synchronized state. This innovative technique is instrumental in facilitating both expansive (global) and targeted (local) synchronization learning processes.

First, DLS is validated using the relatively simple FitzHugh-Nagumo (FHN) nonlinear model, coupled with elementary electrical synapses. Its effects are examined across a range of network distributions including fully connected, small-world, scale-free, and random networks, with all results demonstrating DLS's significant impact. The successful application of DLS in larger, more complex networks underscores its broad applicability. Subsequently, the efficacy of self-adaptive DLS in achieving local synchronization is confirmed, showing its high effectiveness within suitable local ranges. In supervised DLS studies, substantial outcomes are observed in both global and local synchronization contexts. Moreover, the hybrid application of self-adaptive and supervised DLS within the same network is also investigated. The findings indicate that a smaller scope of supervised DLS action leads to desynchronization between the two parts, whereas a larger scope results in overall network synchronization.

Ultimately, the more complex Hodgkin-Huxley (HH) nonlinear model, coupled with chemical synapses, serves to evaluate the effectiveness of DLS. The concluding results demonstrate that DLS significantly facilitates the emergence of synchronized states in highly complex nonlinear networks under suitable conditions. The insights gleaned from this study contribute novel perspectives to the domain of synchronization within coupled nonlinear systems.

**Funding**   This work is supported by National Natural Science Foundation of China under Grants No. 12175080, and also financially supported by self-determined research funds of CCNU from the colleges' basic research and operation of MOE under No. CCNU22JC009.

**Data availability**   The data that support the findings of this study are available from the corresponding author upon reasonable request.

**Declarations**

**Conflict of interest**   The authors declare that they have no potential conflict of interest.